\documentclass[final,3p,times,twocolumn]{elsarticle}
\usepackage{amssymb}

\usepackage[breaklinks=true,colorlinks=true,linkcolor=blue,urlcolor=blue,citecolor=blue]{hyperref}

\usepackage{graphicx}
\usepackage{amssymb}
\usepackage{braket}
\usepackage{slashed}
\usepackage{epstopdf}
\usepackage{mathtools}

\usepackage{subcaption}
\journal{Nuclear Physics B}
\begin{document}
\begin{frontmatter}

\title{Chiral symmetry breaking for fermions charged under large Lie groups}
\author{Felipe J. Llanes-Estrada and Alexandre Salas-Bern\'ardez}
\address{Depto. F\'{\i}sica Te\'orica, Universidad Complutense de Madrid, Plaza de las Ciencias 1, 28040 Madrid, Spain.}

\begin{abstract}
We reexamine the dynamical generation of mass for fermions charged under various Lie groups with equal charge and mass at a high Grand Unification scale, extending the 
Renormalization Group Equations in the perturbative regime to two-loops and matching to the Dyson-Schwinger Equations in the strong coupling regime. 
\end{abstract}

\end{frontmatter}

\section{Introduction}
The Standard Model is a gauge field theory based on the gauged symmetry
\begin{equation}\label{eq:SM}
SU(3)_C\otimes SU(2)_L\otimes U(1)_Y\;. 
\end{equation}
Here $SU(3)_C$ denotes the color interaction responsible for the strong force, $SU(2)_L$ the isospin coupling of left-handed fermions and $U(1)_Y$ the hypercharge group. The spontaneous breaking of the electroweak symmetry by the Higgs mechanism suggested the possibility of higher symmetries at yet higher scales that would also be spontaneously broken, providing strong and electroweak force unification at higher scales; these symmetries would also have to be spontaneously broken~\footnote{It is usually and superficially stated that the gauge symmetry $SU(2)_L\times U(1)_Y$ is spontaneously  broken. However, Elitzur's theorem~\cite{Elitz} states that gauge symmetries cannot be spontaneously broken. First they must be broken explicitly by a gauge fixing term leaving only the global symmetry and then this remaining symmetry can be spontaneously broken. The modern viewpoint is that gauge symmetries are just a redundancy in the description of the theory on which expectation values of observables must not depend. The actual symmetry from which consecuences such as degeneracies in the spectrum, couplings or conserved currents appear is the true global symmetry. We will continue using ``spontaneous symmetry breaking'' without specifying, though in the understanding that it is the global group which is affected.}.
 
In the SM, the Higgs vacuum expectation value breaks the global symmetry $SU(2)_L\times U(1)_Y$ of the Higgs sector in the SM down to $U(1)_{em}$~\cite{SymSM} (or, considering the $U(1)$ as a perturbation, and the approximate global custodial $SU(2)$, it breaks $SU(2)\times SU(2)\to SU(2)_c$). This generates masses for the $W^\pm$ and $Z$ bosons, and for fermions, leaving us the symmetry
\begin{equation}
SU(3)_C\otimes U(1)_{em}\;
\end{equation}
(and the approximate custodial $SU(2)$).
A feature of the symmetry group of the Standard Model that stands out is the small size of the numbers 1-2-3. Why are we confronted by such symmetry groups? Why not larger groups like $SU(6)$ or $Sp(10)$? 

To address these question we study in Section~\ref{FromGUT} how hypothetical quarks colored under different groups acquire masses from a Grand Unified Theory (GUT) scale where all groups under consideration are chosen to have the same couplings and quark masses, down to lower energies where the interaction becomes strong. For this task we will use the Renormalization Group Equations (RGE).

Then, section~\ref{sec:DSE} treats the Dyson-Schwinger Equations (DSE) for the lowest scales when the interactions become strong. 
Any workable truncation of the DSE typically fails to satisfy local gauge invariance, while respecting global symmetry. This is however enough to discuss its breaking in view of Elitzur's theorem. While realistic models~\cite{GG} that embed the SM such as $SU(5)$ or $SO(10)$ are often discussed~\footnote{While the absence of proton decay rules out some classic implementations of the GUT idea, models keep being constructed that evade the constraints~\cite{Fornal:2018aqc}}, we are here less ambitious and keep the discussion at a general level, considering multiple groups.

In addition to a brief discussion in section~\ref{sec:outlook}, the article has an appendix addressing the computation of color factors for almost all of the continuous Lie groups (results for E8 are not at hand). We have kept the article as short as is compatible with its being self-contained, since the theory behind our approach has already been laid out in a previous publication~\cite{Llanes}. We have striven to extend that calculation as explained next.

\section{From Grand Unification to strong interaction scale with the Renormalization Group Equations}\label{FromGUT}
Our motivation in this work is to extend the one-loop RGE computation of \cite{Llanes} to two loops. This was the aspect that introduced the most uncertainty to predict the mass of the fermions charged under large groups. In doing so we have unveiled partial errors in the original publication that we here correct. An erratum has also been issued to warn the reader of the earlier article.

We evolve the masses of one single color-charged fermion for the different color groups from the Grand Unification scale of $\mu_{GUT}=10^{15}\;GeV$ to the point where interactions become strong (at a scale $\sigma$) for each group, that is, when $C_F\,\alpha_s(\sigma)=0.4$. Once this happens we use Dyson-Schwinger equations for the non-perturbative regime in order to obtain the constituent masses for these fermions:  this step is explained in the next section~\ref{sec:DSE}.

An efficient way of keeping track of the parameter evolution needed for the physical predictions of a theory to be invariant under $\mu$ scale choice is the use of RGEs. We generalize those of Quantum Chromodynamics to an arbitrary gauge group $G$. The running of the coupling constant $g_s$ with  $\mu$ is~\cite{TMuta} determined by the $\beta(g_s)$ function,
\begin{equation}
\beta(g_s)\equiv -\mu\frac{dg_s}{d\mu}=\beta_1g_s^3+\beta_2 g_s^5+...\;.
\end{equation}
 The one-loop correction $\beta_1$ is
\begin{equation}
\beta_1=\frac{1}{(4\pi)^2}\left(\frac{11C_G-2T_RN_f}{3}\right)\;,
\end{equation}
where $C_G$ is the adjoint Casimir~\footnote{For $SU(N)$, $C_G=N$, the group dimension. But in general, $C_G=aN+b$ with $a,b$ depending on the particular group, as listed in the appendix. This detail was in error in~\cite{Llanes} and is being corrected.}, $T_R$ the normalization of the generators $T^a$ of the group $G$ defined as $Tr(T^aT^b)\equiv T_R\delta^{ab}$ and $N_f$ the number of colored fermions~\footnote{In this article we take $N_f=1$, but a brief discussion in~\cite{Llanes} reminds us that there is a critical number of colors $N_f^c$ that shuts off the vacuum antiscreening and thwarts spontaneous symmetry breaking.}.

\begin{figure}[!ht]
\includegraphics[width=1\columnwidth]{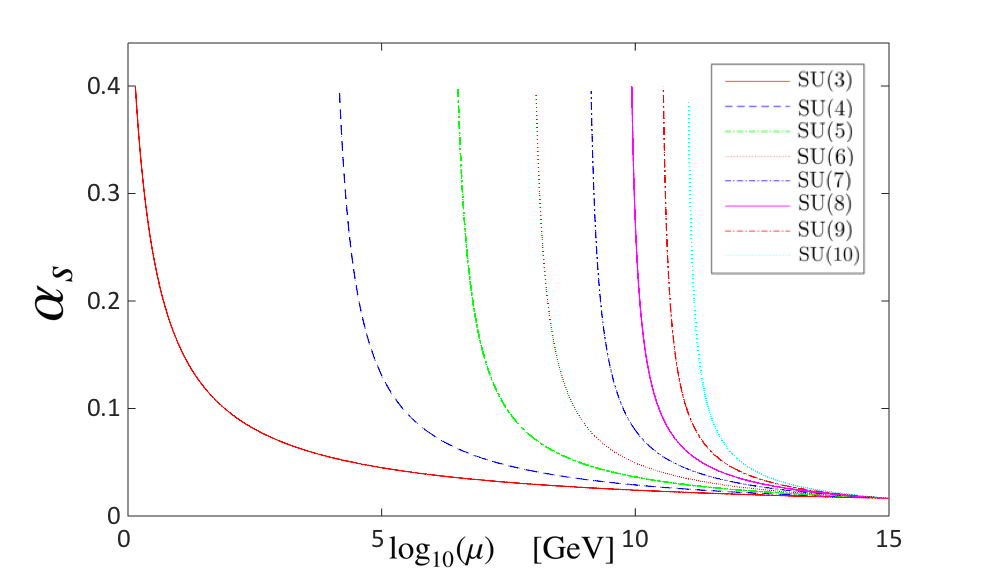}
\includegraphics[width=1\columnwidth]{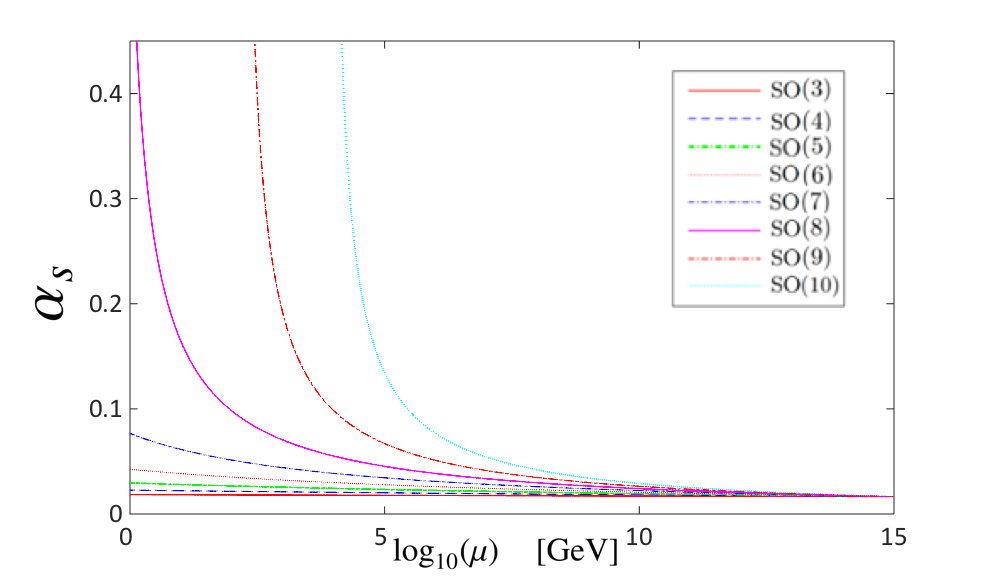}
\includegraphics[width=1\columnwidth]{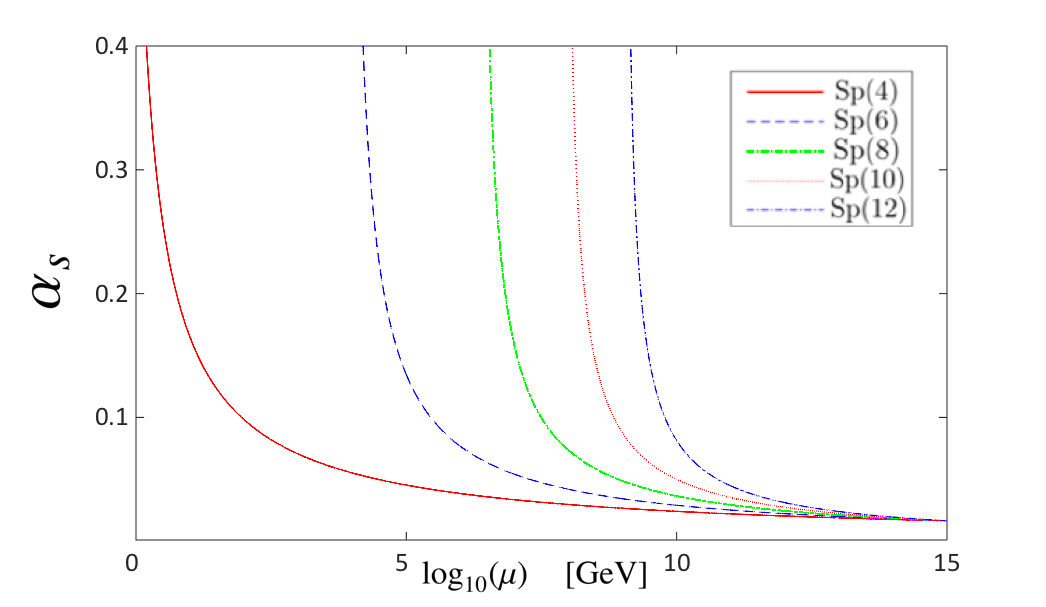}
\includegraphics[width=1\columnwidth]{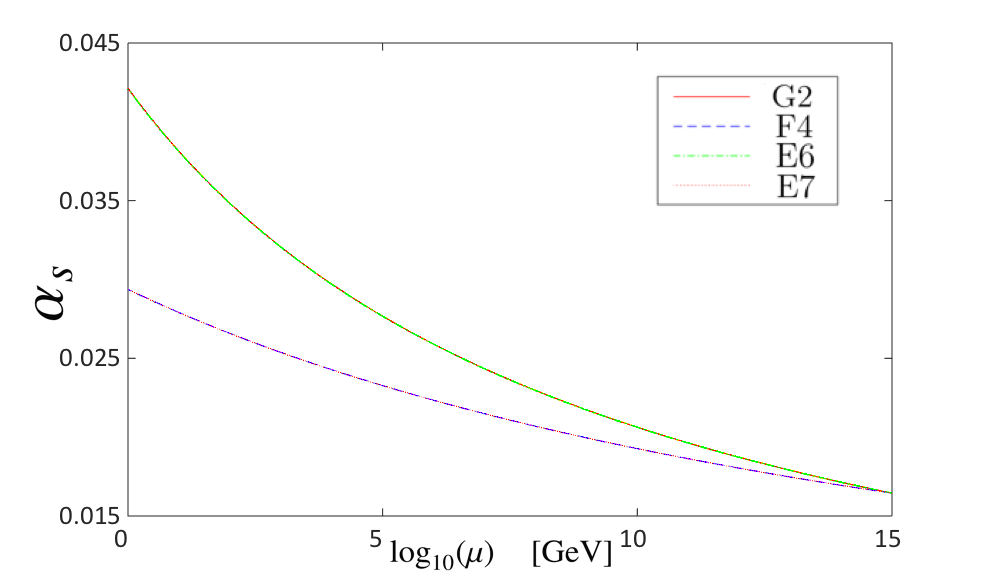}
  \caption{Running couplings for the families $SU(N)$, $SO(N)$, $Sp(N)$ and most of the Exceptional Lie groups. \label{fig:runningcoupling}}
\end{figure}

The two loop contribution to the $\beta(g_s)$ function, $\beta_2$, entails a larger effort in perturbation theory, but can also be easily found in the literature~\cite{TMuta}, 
\begin{equation}
\beta_2=\frac{1}{(4\pi)^4}\left(\frac{34}{3}C_G^2-4\left(\frac{5}{3}C_G+C_F\right)T_RN_f\right)\;,
\end{equation}
where $C_F$ is the Casimir of the fundamental representation(see appendix).
Using the color coefficients listed there, we obtain the running couplings of
$SU(N)$, $SO(N)$, $Sp(N)$ and the exceptional groups $G2$, $F4$, $E6$ and $E7$, shown in Figure~\ref{fig:runningcoupling}.

\newpage

The result of~\cite{Llanes}, that for small groups and one flavor $\sigma \propto e^{N}$ stands out. The very large groups have strongly interacting scales $\sigma$ clustering around the GUT scale, since they run very fast. We are then ready to start employing the DSEs down from the scale $\sigma$.

Simultaneously, running of the current mass $m_c$ is set by the self energy correction to the quark propagator that implies an anomalous mass dimension $\gamma_m$
\begin{equation}
\gamma_m(g_s)\equiv -\frac{\mu}{m}\frac{dm}{d\mu}=\gamma_1g_s^2+\gamma_2 g_s^4+...\;.
\end{equation}
The one loop contribution to the $\gamma_2(g_s)$ function for the quarks, $\gamma_1$, amounts to
\begin{equation}
\gamma_1=\frac{6C_F}{(4\pi)^2}\;.
\end{equation}

The two loop contribution to $\gamma_m(g_s)$ (see \cite{Tarrach}), $\gamma_2$, is
\begin{equation}
\gamma_2=\frac{C_F}{(4\pi)^4}\Big(3C_F+\frac{97}{3}C_G-\frac{20}{3}T_R N_f\Big)\label{eq:gamma2}\;.
\end{equation}

\begin{figure}[!ht]
\includegraphics[width=1\columnwidth]{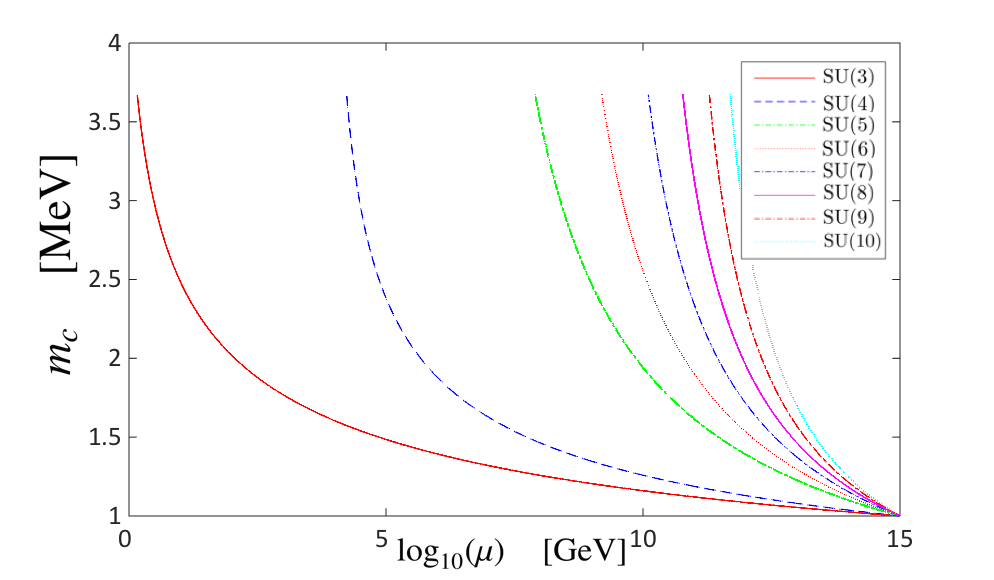}
\includegraphics[width=1\columnwidth]{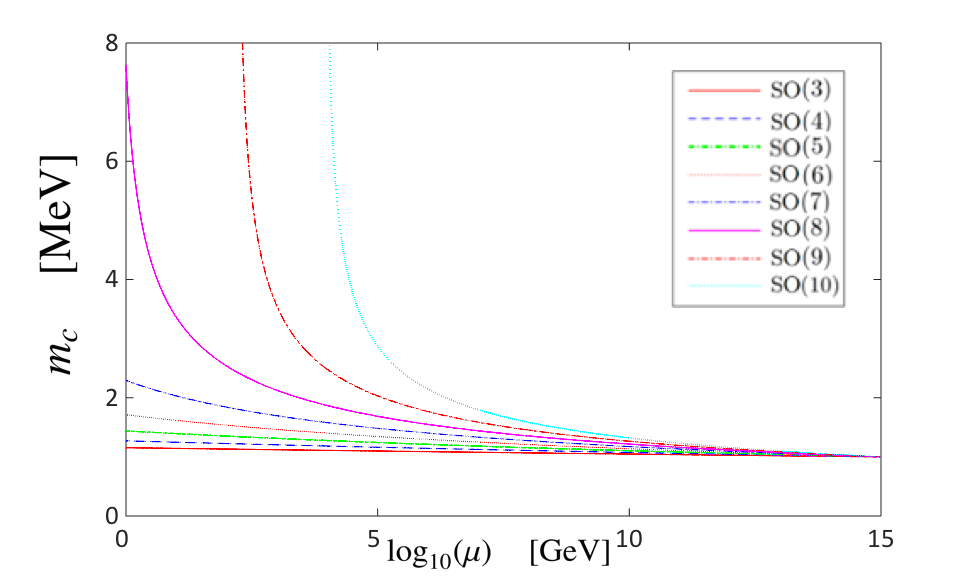}
\includegraphics[width=1\columnwidth]{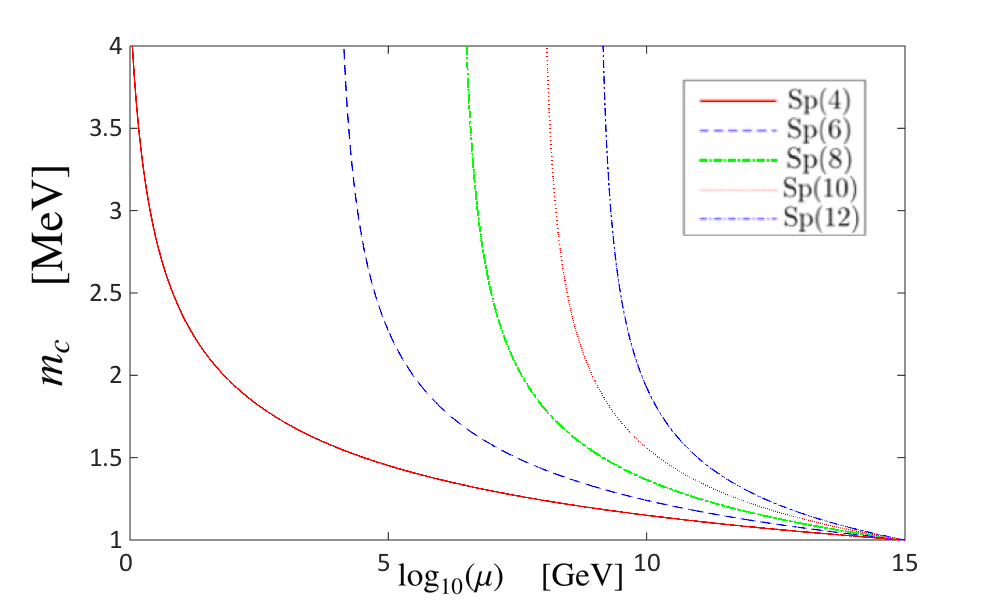}
\includegraphics[width=1\columnwidth]{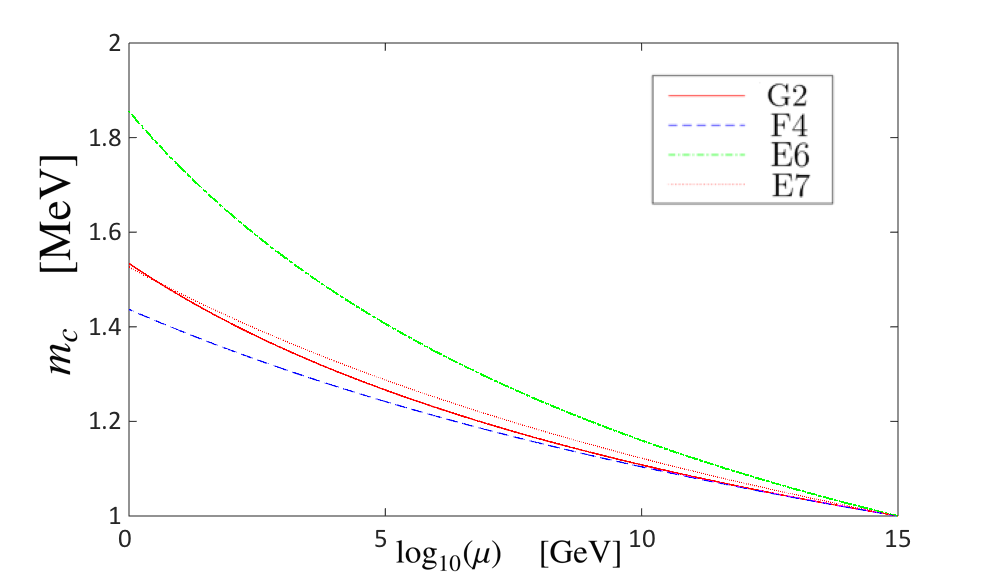}
  \caption{Running masses for the families $SU(N)$, $SO(N)$, $Sp(N)$ and four out of five Exceptional Lie groups.\label{fig:runningmass}}
\end{figure}

At the GUT  starting scale of the RGEs we choose a fermion mass $m_{c}(\mu_{GUT})=1\;MeV$ and fix the coupling $\alpha_{s}(\mu_{GUT})\equiv g_s({\mu_{GUT}})^2/4\pi=0.0165$ to broadly reproduce the isospin average mass for the $SU(3)_C$ quarks of the first generation at the scale $\mu=2\;GeV$,
\begin{equation}
\overline{m}(2\,GeV)=\frac{m_u(2\;GeV)+m_d(2\;GeV)}{2}\simeq3.5\;\;MeV\;.
\end{equation}
These initial conditions are taken to be the same for all Lie groups, as suggested by the concept of GUT. Then, the mass running for the various Lie groups, with color factors taken from \ref{sec:CF} is plotted in Figure~\ref{fig:runningmass}.\\

\newpage

$ $

\newpage
\section{Running at the strong interaction scale with the Dyson-Schwinger Equations}
\label{sec:DSE}

Once the interactions become strong, perturbation theory breaks down and resummation becomes necessary: we thus adopt the simplest possible DSE for the quark self energy. The free propagator of a fermion with current mass $m_c$~\cite{TMuta}, $ S_0(p^2)=\frac{1}{m_c-\slashed p}\;,$
 becomes a fully dressed one $ \tilde{S}(p^2)=\frac{1}{B(p^2)-A(p^2)\slashed p}\;$. 
Being only interested in qualitative features of spontaneous mass generation, we can approximate $A(p^2)=1$ which leaves the physical mass as $M(p^2)\equiv B(p^2)$.
Denoting $\Sigma(p)$ as the sum of all one-particle irreducible diagrams, the DSE takes the form (omitting the $p$ dependence)
\begin{equation}
\tilde{S}(p^2)=S_0(p^2)\,(1-\Sigma(p) S_0(p^2))^{-1}\;.
\end{equation}
 Inverting, we see that
$ \tilde{S}^{-1}(p^2)=S_0(p^2)^{-1}-\Sigma(p) \Rightarrow M(p^2)=m_c-\Sigma(p)\;. $

\begin{figure}[!ht]\centering
\includegraphics[width=0.9\columnwidth]{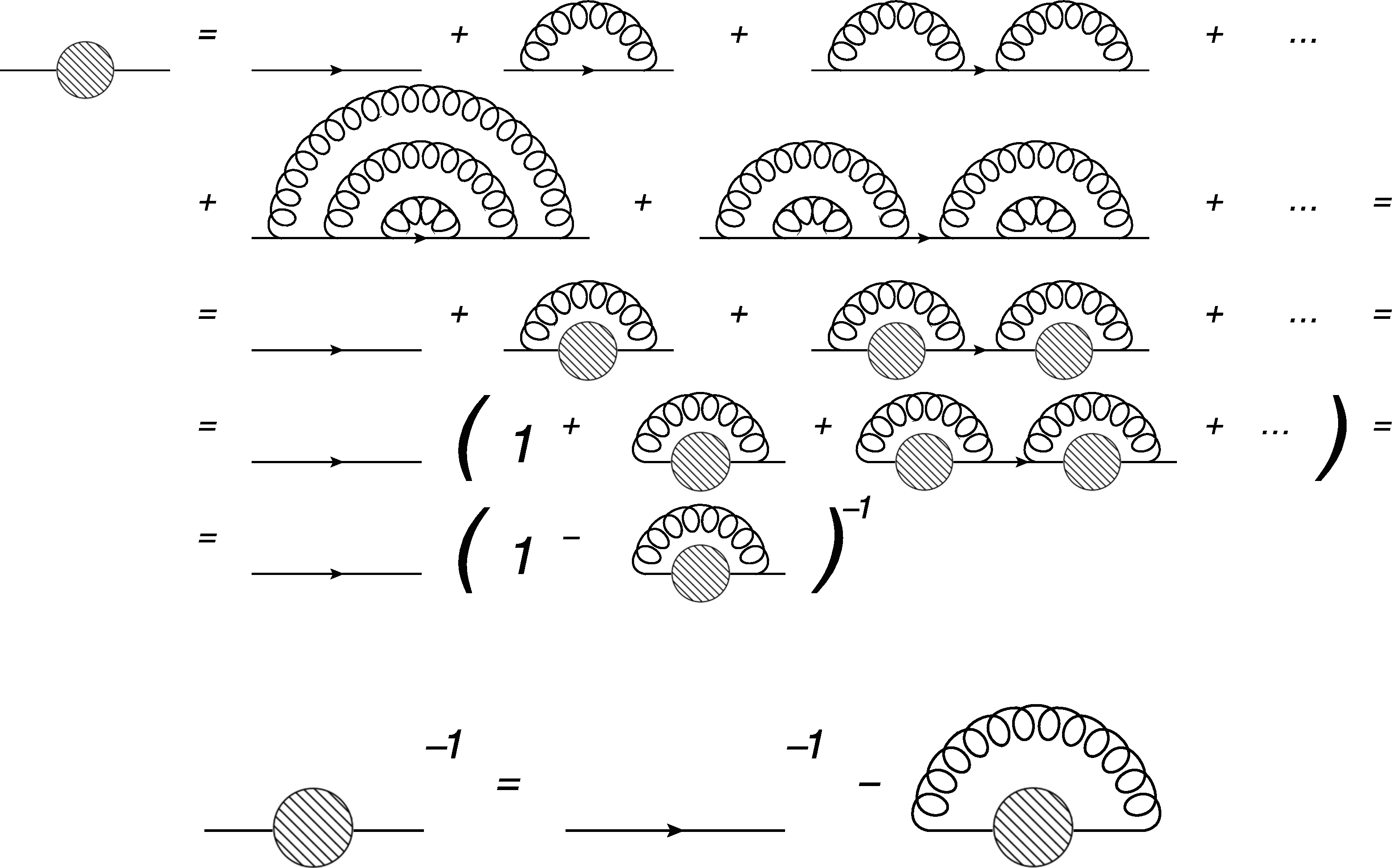}\\
\vspace{1cm}
\includegraphics[width=0.7\columnwidth]{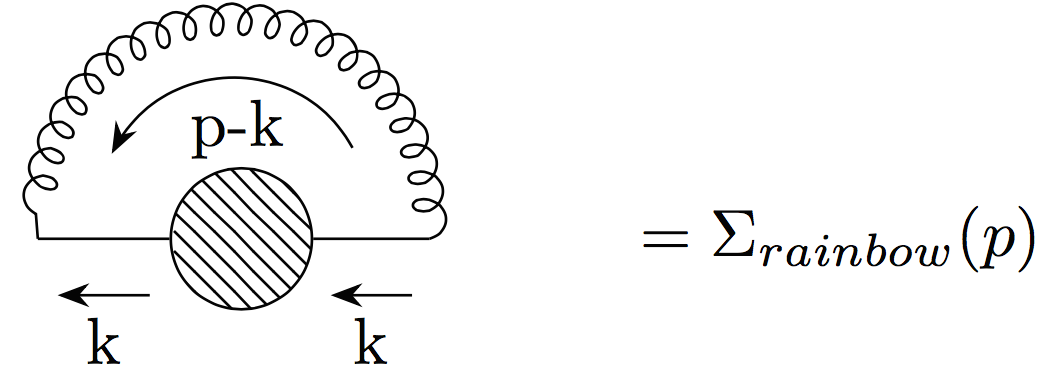}
  \caption{Rainbow DSE for the full quark propagator (filled circle).\label{fig:rainbow}}
\end{figure}

To illustrate the possibilities, we will employ the \textit{rainbow truncation} that sums only ``rainbow shaped'' diagrams, with great simplification (Fig.~\ref{fig:rainbow}).
\\
 The one-loop self energy is then, passing to Euclidean space with $k^0\to ik^0$, 
$p^0\to ip^0$, given by
\begin{align}
\Sigma_{\rm rainbow}(p)=&g^2_s\int\frac{d^4k}{i(2\pi)^4}\gamma^\mu(T^a)\frac{1}{M(k^2)-\slashed k}\gamma^\nu(T^a) \frac{\eta_{\mu\nu}}{(k-p)^2}\nonumber\\
= C_F&g^2_s\int_0^\infty\frac{dk_E \,k_E^3}{\pi^3}\frac{-M(k^2)}{M^2(k^2)+k_E^2}\nonumber\times\\
\times&\int_{-1}^{+1}\sqrt{1-x^2}\frac{dx}{(k_E^2-2|k_E||p_E|\,x+p^2_E)}\;.
\end{align}
We define the last integral in $x$ as the averaged gluon propagator $D^0_{k-p}$ (in the Feynman Gauge) over the four dimensional polar angle. Hence, we conclude that the Dyson-Schwinger equation in the rainbow approximation for the quark propagator is
\begin{equation}\label{eq:DSE}
M(p^2)=m_c+C_Fg^2_s\int_0^\infty\frac{dq \,q^3}{\pi^3}\frac{M(q^2)}{M^2(q^2)+q^2}D^0_{q-p}\;.
\end{equation}

Note that the integral in (\ref{eq:DSE}) is divergent and must be regularized. We could employ a simple cutoff regularization cutting this integral at a scale $\Lambda$; instead we would like to preserve Lorentz invariance and exhibit renormalizability. Following again \cite{Llanes}, we introduce renormalization constants $Z(\Lambda^2,\mu^2)$ to absorb infinities and any dependence on the cutoff $\Lambda$,
\begin{equation}
\tilde{S}^{-1}(p^2,\mu^2)\equiv Z_2 S_0^{-1}(p^2)-\Sigma(p^2,\mu^2)\;,
\end{equation}
where the dependence of $\Sigma$ on $\mu$ is given by the fermion and gluon propagators. Apart from the wavefunction renormalization $Z_2$ we introduce $Z_m$ for the bare quark mass. The relation between the (cutoff dependent) unrenormalized mass $m_c(\Lambda^2)$ and the renormalized mass at the renormalization scale $\mu$, $m_R(\mu^2)$, is
\begin{equation}
m_c(\Lambda^2)=Z_m(\Lambda^2,\mu^2)m_R(\mu^2).
\end{equation}
Since we will maintain the restriction $A(p^2)=1$, renormalization of the quark wavefunction is not necessary, therefore $Z_2=1$. The only renormalization condition is to fix the  mass function at $p^2=\mu^2$. The DSE is then
\begin{equation}
M(p^2)=Z_m m_R(\mu^2)-\Sigma(p^2,\mu^2)\;.\label{EQ:RMM}
\end{equation}
Evaluating (\ref{EQ:RMM}) at $p^2=\mu^2$ and subtracting it again to (\ref{EQ:RMM}) we obtain,
\begin{align}\label{eq:DSEMOM}
M(p^2)&=M(\mu^2)\nonumber\\
&+C_Fg^2_s\int_0^\infty\frac{dq \,q^3}{\pi^3}\frac{M(q^2)}{M^2(q^2)+q^2}\Big(D^0_{q-p}-D^0_{q-\mu}\Big)\;.
\end{align}
It is easy to show, taking $\mu$ parallel to $p$, that asymptotically~\cite{Llanes},
\begin{equation}
\frac{\partial M(p^2)}{\partial \Lambda}\propto\frac{M(\Lambda^2)(p-\mu)}{\Lambda^2}\;.
\end{equation}
Therefore, for large $\Lambda$, $M(p^2)$ stops depending on the cutoff, which can be taken {\it e.g.} to $\Lambda=10^{10}\; GeV$ and renormalization is achieved.

\begin{figure}[!h]
\includegraphics[width=1\columnwidth]{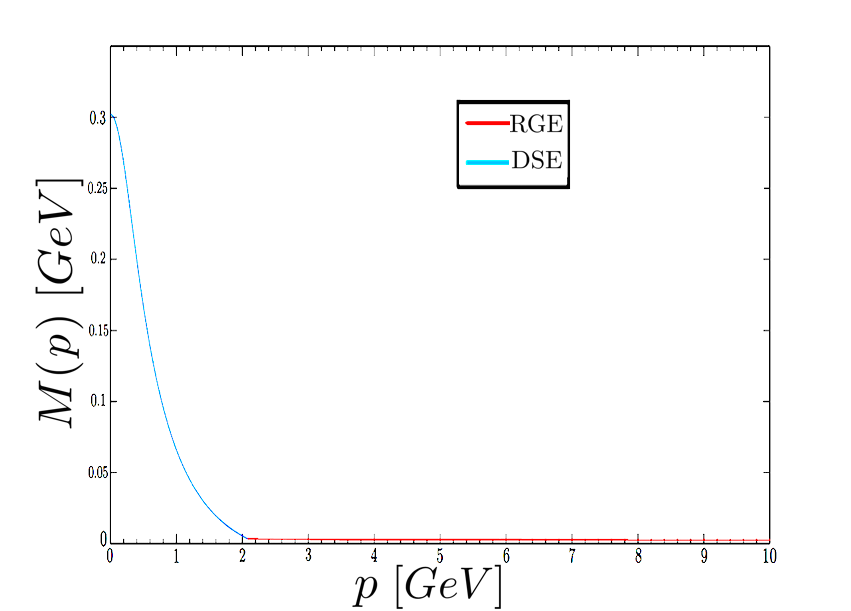}
\caption{Matching of RGE and DSE solutions of the Mass Running for $SU(3)$.}\label{fg:DSERGE}
\end{figure}

Now we are ready to obtain the quark constituent masses for all the groups studied. We match the RGE solution (high scales) to the DSE solution (low scales) at the matching energy $\sigma$ where interactions become strong,  $C_F\alpha_s(\sigma)=0.4$ for each group, as advertised. 
For SU(3) ($C_F=\frac{4}{3}$), the scale where $\alpha_s(\sigma)=0.3$ is $\sigma=2.09\; GeV$. From this point down in scale we freeze $\alpha_s$.  A constant vertex factor of order 7 is applied to the DSE to guarantee sufficient chiral symmetry breaking at low scales, requiring the constituent quark mass $M(0)$ to be close to $300\;MeV$ using the substracted DSE (\ref{eq:DSEMOM}). This is supposed to mock up the effect of vertex-corrections not included, and is known to scale with $N$~\cite{Alkofer:2008tt} for large $N$, the group's fundamental dimension.
Finally, the $M(p)$ obtained is plotted in Figure \ref{fg:DSERGE}.

\begin{figure}[!ht]
\includegraphics[width=1.1\columnwidth]{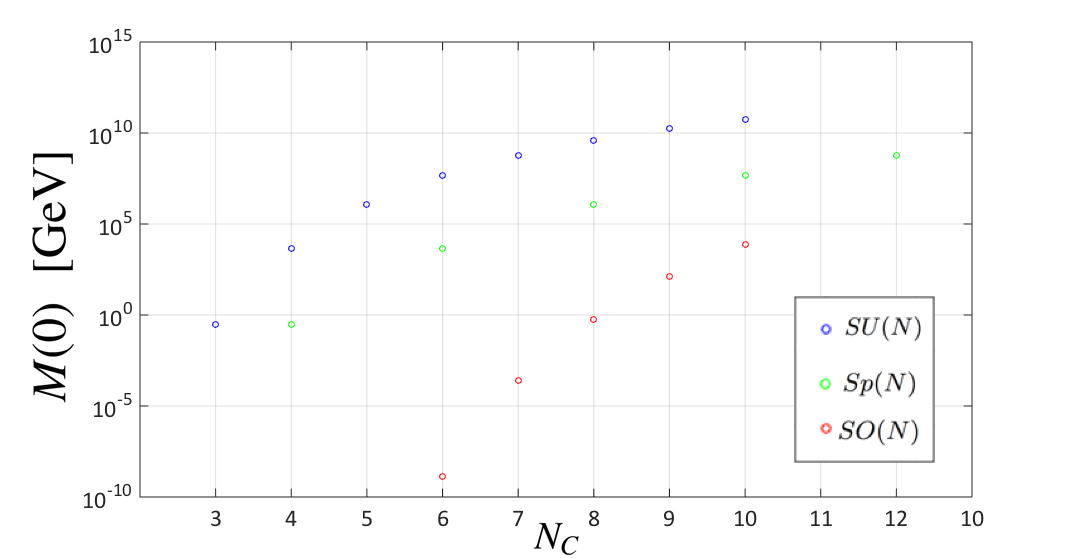}
\caption{Constituent Masses for the groups which break chiral symmetry in RGE before $10^{-5}\;GeV$.}\label{constituent}
\end{figure}

To obtain the constituent fermion masses for the different Lie Groups we use a trick presented in \cite{Llanes}: to perform a scale transformation 
\begin{equation}
p^2\to \lambda^2p^2 \ ; \ \sigma^2\to \lambda^2\sigma^2
\end{equation}
on the DSE (\ref{eq:DSEMOM}). Changing the integration variable $q^2\to \lambda^2q^2$, giving $d^4q\to \lambda^4d^4q$, 
the modified DSE equation is satisfied by a modified $\tilde{M}$  and the relation between the constituent masses is simply $ M(0)=\frac{\tilde{M}(0)}{\lambda}\;.$
Now, taking $\lambda$ as the ratio of the scales where interactions become strong for $SU(3)$ and another group,
\begin{equation}
\frac{\sigma_{group}}{\sigma_{SU(3)}}=\lambda\;,
\end{equation}
the mass function scales in the same way,
\begin{equation}
\frac{M_{group}(0)}{M_{SU(3)}(0)}=\lambda\;.
\end{equation}
Hence, eliminating the auxiliary $\lambda$, we find
\begin{equation}
\frac{M_{group}(0)}{M_{SU(3)}(0)}=\frac{\sigma_{group}}{\sigma_{SU(3)}}\,.
\end{equation}
Using these results we compute the constituent masses for the quarks charged under the different groups (Fig. \ref{constituent}). 
The outcome is that the special Lie groups examined do not spontaneously generate fermion mass at a high scale: their interactions, running at two loops from the GUT, are too weak to do so. This is because the $C_G$ Casimir of the adjoint representation, though proportional to the group dimension, carries a small numerical factor that reduces the intensity of coupling running.

Large $SU(N)$ and $Sp(N)$ groups, on the other hand, behave as advanced in~\cite{Llanes}, and generate a mass for the fermions that puts them beyond reach of past accelerators. The exceptions are $Sp(4)$, for which the mass generation is similar to QCD; and $Sp(2)$ , which is too weak.  
As for the special orthogonal groups, for $SO(N>10)$, once more the fermion mass generated is too large to be accessible at accelerators.

\section{Conclusions and Outlook}\label{sec:outlook}

We have examined mass generation for different Lie groups with an arbitrary number of colours. As a definite starting point, we have adopted the philosophy of Grand Unification in which fermion masses as well as coupling constants, for all groups, coincide at a high scale, namely $10^{15}$ GeV.

We have run the couplings and masses for each group to lower scales employing two-loop Renormalization Group Equations, using as an input the Cuadratic Casimirs obtained in~\ref{sec:CF}.
We chose the initial conditions at $\mu=\mu_{\rm GUT}$ to be the same for all groups and selected so that $SU(3)_C$ at the scale of 2 GeV yields a rough approximation of the strong force coupling and first-generation isospin-averaged quark mass. 

Typically, for all but the smallest groups, a scale arises where interactions become strong (discernable as a Landau pole in perturbation theory). We stop running at the scale $\mu$ such that $\alpha_s(\mu)C_F=0.4$; below that, we employ a non-perturbative treatment, namely Dyson-Schwinger Equations in the rainbow approximation to assess the masses down to yet lower scales.

Combining the methods of RGE and DSE and requiring that the constituent masses of $SU(3)_C$ colored quarks to be 300 MeV has allowed us to obtain the constituent masses of hypothetical fermions charged under different groups from a Grand Unification Scale of $10^{15}\;\text{GeV}$. 
From this treatment we can conclude that groups belonging to the $SU(N)$ and $Sp(N)$ families, with $N>4$, generate masses of order or above the few TeV. Notwithstanding the crude approximations we have employed, our computation gives about 5 TeV to $SU(4)$-charged fermions, which would not be far out of reach of mid-future experiments provided the GUT conditions apply. It appears from our simple work that larger groups (except the Exceptional Groups and $SO(N)$ with $N<10$) might endow fermions with a mass too high to make them detectable in the foreseeable future.
In case these superheavy fermions would have been coupled to the Standard Model, they would have long decayed in the early universe due to the enormous phase space available. If they existed and be decoupled from the SM, they would appear to be some form of dark matter.  We have also provided a partial answer to the question ``\textit{Why the symmetry group of the Standard Model, $S U (3)_C \otimes SU(2)_L \otimes U(1)_Y$ , contains only small-dimensional subgroups?}'' It happens that, upon equal conditions at a large Grand Unification scale, large-dimensioned groups in the classical $SO(N)$, $SU(N)$ and $Sp(N)$ families force dynamical mass generation at higher scales because their coupling runs faster. Should fermions charged under these groups exist, they would appear in the spectrum at much higher energies than hitherto explored \cite{Odense}. \\

Interestingly for collider phenomenology, we find the masses of fermions charged under the following groups are within reach of the energy frontier: 
$M_{\rm SU(4)}\simeq 5$ TeV; $M_{Sp(6)}\simeq4.4$ TeV; $M_{\rm SO(10)}\simeq 7$ TeV.
The LHC might be able to exclude those~\footnote{For comparison, the one-loop results are 
$M_{\rm SU(4)}\simeq 2$ TeV; $M_{Sp(6)}\simeq1.5$ TeV; $M_{\rm SO(10)}\simeq 3$ TeV., which indicates fair convergence.}.

However, the following groups $SO(N<10)$, $E_6$, $E_7$, $G2$ and $F4$ yield masses that are below the TeV scale and should already have been seen if they coupled to the rest of the Standard Model (one could argue that those isomorphic to groups present in the SM have already been sighted). Their absence from phenomenology thus suggests that fermions charged under any of those groups , if at all existing, belong to a decoupled dark sector. 

\appendix
\section{Color Factors} \label{sec:CF}
We here present some of the calculations carried out to obtain the quadratic Casimirs needed in both RGE and DSE. Such quadratic Casimirs are elements in the Lie Algebra which commute with all the other elements (See~\cite{Cvit,GTN,FFS} for the necessary group theory). \\
We will focus on the Casimir invariant in the fundamental representation of the group $G$, $C_F\delta^{ij}=(T^a T^a)^{ij}$, and the Casimir invariant in the adjoint representation, $C_G\delta^{ab}=f^{acd}f^{bcd}$. Normalization of the algebra generators is chosen as $Tr(T^aT^b)=\frac{1}{2}\delta^{ab}$.
\subsection{Special Unitary Groups $SU(N)$}
We start with the special unitary family $SU(N)$. Its generators $T^{a}$ are traceless hermitian. Therefore every Hermitean $N\times N$ matrix $A$ can be written as,
\begin{equation}
A=A^\dagger=c_0\mathbb{I}+c_a T^a\,.
\end{equation}
From this we find
\begin{equation}
c_0=\frac{Tr(A)}{N}\;\;\;\;\;\;c_a=2\, Tr(AT^a)\,.
\end{equation}
Having then
\begin{align}
&A_{ij}=A_{lk}\delta^{li}\delta^{kj}=A_{lk}\Big(2(T^a)_{ij}(T^a)^{kl}+\frac{1}{N}\delta^{kl}\delta_{ij}\Big)\\
&\Rightarrow A_{lk}\Big(2(T^a)_{ij}(T^a)^{kl}+\frac{1}{N}\delta^{kl}\delta_{ij}-\delta^{li}\delta^{kj}\Big)=0\;.
\end{align}
Since $A$ is arbitrary, we find for the generators the useful relation 
\begin{equation}
(T^a)_{ij}(T^a)_{kl}=\frac{1}{2}\Big(\delta_{li}\delta_{kj}-\frac{1}{N}\delta_{kl}\delta_{ij}\Big)\;.
\end{equation}
Contracting $j$ and $k$ we obtain the fundamental representation Casimir or Color Factor
\begin{equation}
(T^aT^a)_{ij}=\frac{1}{2}\Big(\frac{N^2-1}{N}\Big)\delta_{ij}=C_F\delta_{ij}\;.
\end{equation}
Now we compute the following combination, 
\begin{align}
(T^a)_{i}^j(T^b)_{jk}(T^a)^k_{l}=\frac{1}{2}\Big((T^b)_{jk}\delta_{li}\delta^{kj}-\frac{1}{N}(T^b)_{jk}\delta^k_{l}\delta_{i}^j\Big)\nonumber\\
=-\frac{1}{2N}(T^b)_{il}\;.
\end{align}
 Noting the following identity and using the results already computed, we obtain the adjoint Casimir for $SU(N)$,
\begin{align}\label{eq:CG}
f^{acd}f^{bcd}&=-2\,Tr\Big([T^a,T^c][T^b,T^c]\Big)\nonumber\\
&=-2 Tr\Big(2T^aT^cT^bT^c-(T^aT^b+T^bT^a)T^cT^c\Big)\nonumber\\
&=N\,\delta^{ab}=C_G\delta^{ab}\,.
\end{align}
\subsection{Special Orthogonal Groups $SO(N)$}
We will follow now the same steps for the Special Orthogonal family $SO(N)$. Its generators are antisymmetric and traceless and they form a basis for the antisymmetric $N\times N$ matrices. Thus, taking an antisymmetric matrix $A$, we have
\begin{equation}
A=-A^T=c_a T^a\;\;\;\; \Rightarrow\;\;\;\; c_a=2 Tr\Big(A T^a\Big)\;.
\end{equation}
Then we have
\begin{equation}
A_{ij}=A_{kl} \delta^k_i\delta^l_j=\frac{1}{2}A_{kl}\Big( \delta^k_i\delta^l_j-\delta^k_j\delta^l_i\Big)=A_{lk}\Big(2 (T^a)_{ij}(T^a)^{kl}\Big)\;.
\end{equation}
Finding 
\begin{equation}
A_{kl}\Big[ \frac{1}{2}\Big( \delta^k_i\delta^l_j-\delta^k_j\delta^l_i\Big)+\Big(2 (T^a)_{ij}(T^a)^{kl}\Big)\Big]=0\;,
\end{equation}
and since A is an arbitrary antisymmetric matrix we get
\begin{equation}
(T^a)_{ij}(T^a)^{kl}=\frac{1}{4}\Big(\delta^k_j\delta^l_i- \delta^k_i\delta^l_j\Big)\;.
\end{equation}
Here, since the group is real there is no need for distinction between upper and lower indices. Contracting in the previous expression $j$ with $k$ we obtain the Color Factor
\begin{equation}
(T^a)_{ij}(T^a)^{jl}=\frac{1}{4}\Big(\delta^j_j\delta^l_i- \delta^j_i\delta^l_j\Big)=\frac{N-1}{4}\delta_i^l=C_F\delta_i^l\;.
\end{equation}
As before, we compute
\begin{align}
(T^a)_{ij}(T^b)^{jk}(T^a)_{kl}&=\frac{1}{4}\Big((T^b)^{jk}\delta_{il}\delta_{kj}-(T^b)^{jk}\delta_{ik}\delta_{lj}\Big)\nonumber\\
&=-\frac{1}{4}(T^b)_{li}=\frac{1}{4}(T^b)_{il}\;.
\end{align}
We are able now to obtain the adjoint Casimir for $SO(N)$. Similar to (\ref{eq:CG})
\begin{align}
f^{acd}f^{bcd}&=2\,Tr\Big(\frac{1}{2}T^aT^b-(T^aT^b+T^bT^a)\frac{N-1}{4}\Big)\nonumber\\
&=\frac{1}{2}(N-2)\delta^{ab}=C_G\delta^{ab}\,.
\end{align}

\subsection{Simplectic Groups $Sp(N)$}

The elements $M\in Sp(N)$ (with $N$ even) are $N\times N$ matrices which preserve the antisymmetric tensor 
\begin{equation}
\Omega=
\begin{pmatrix}
0 & \mathbb{I}_{\frac{N}{2}\times\frac{N}{2}}\\
 -\mathbb{I}_{\small{\frac{N}{2}\times\frac{N}{2}}} & 0
\end{pmatrix},
\end{equation}
in the sense
\begin{equation}
\Omega=M^T\Omega \,M\Rightarrow\;\;M^{-1}=\Omega^TM^T\Omega\,.
\end{equation}
Using this relation it is possible to prove that the generators of the group take the form
\begin{equation}\label{eq:Spgen}
-T^a=\Omega^T(T^a)^T\Omega\;\;\Rightarrow\;\;T^a=
\begin{pmatrix}
A & B\\
 C & -A^T
\end{pmatrix},
\end{equation}
where $B$ and $C$ are symmetric matrices. It is now possible to show that the generators  satisfy
\begin{equation}
(T^a)_{ij}(T^a)_{kl}=\frac{1}{4}\Big(\delta_{il}\delta_{jk}+\Omega_{ik}(\Omega^{-1})_{jl}\Big)\;.
\end{equation}
Therefore
\begin{equation}
(T^a)_{ij}(T^a)^j_{l}=\frac{1}{4}(N+1)\delta_{ij}=C_F\delta_{ij}\;.
\end{equation}
Noticing $\Omega^T=\Omega^{-1}=-\Omega$, we compute the usual combination
\begin{align}
(T^a)_{ij}(T^b)^{jk}(T^a)_{kl}&=\frac{1}{4}\Big(\delta_{il}(T^b)^{jk}\delta_{jk}+\Omega_{ik}(T^b)^{jk}(\Omega^{T})_{jl}\Big)\nonumber\\
&=\frac{1}{4}(\Omega^{T})_{ik}((T^b)^T)^{kj}\Omega_{jl}=-\frac{1}{4}(T^b)_{il}\;,
\end{align}
where in the last equality we have used (\ref{eq:Spgen}). 
The adjoint Casimir now falls down easily
\begin{align}
f^{acd}f^{bcd}&=-2\,Tr\Big(-\frac{1}{2}T^aT^b-(T^aT^b+T^bT^a)\frac{N+1}{4}\Big)\nonumber\\
&=\frac{1}{2}(N+2)\delta^{ab}=C_G\delta^{ab}
\end{align}
To obtain the Color Factors and adjoint Casimirs for $G2$, $F4$, $E6$ and $E7$ we refer to the article of P. Cvitanovi\'c \cite{Cvit}. The results obtained are presented in Table I.\\
\begin{table*}[!ht]
  \begin{tabular}{ | l || c | c | c|}
   \hline
    Group & Color Factor $(C_F)$ & Adjoint Casimir $(C_G)$ & $N_c$\\ \hline\hline
     $SU(N)$ & $\frac{1}{2}\Big(N-\frac{1}{N}\Big)$ & $N$ & $\forall N\in\mathbb{N}$\\ \hline
    $SO(N)$ & $\frac{1}{4}\Big(N-1\Big)$ & $\frac{1}{2}\Big(N-2\Big)$& $\forall N\in\mathbb{N}$ \\\hline
    $Sp(N)$ & $\frac{1}{4}\Big(N+1\Big)$  & $\frac{1}{2}\Big(N+2\Big)$& $N=2n$ $\forall n\in\mathbb{N}$ \\\hline
    $G2$ & $\frac{1}{4}\Big(N-3\Big)$ & $\frac{1}{2}\Big(N-3\Big)$ & $N=7$\\\hline
    $F4$ & $\frac{1}{18}\Big(N-8\Big)$ & $\frac{1}{18}\Big(N+1\Big)$& $N=26$ \\\hline
    $E6$ &  $\frac{1}{12}\Big(N-\frac{29}{3}\Big)$ & $\frac{1}{12}\Big(N-3\Big)$& $N=27$ \\\hline
    $E7$ & $\frac{1}{48}\Big(N+1\Big)$& $\frac{1}{48}\Big(N+16\Big)$  & $N=56$\\
    \hline
  \end{tabular}
  \caption{Cuadratic Casimirs for different Lie Groups.}
\end{table*}


\end{document}